\documentclass[reprint,amsmath,amssymb,aps]{revtex4-1}

\usepackage{graphicx}
\usepackage{dcolumn}
\usepackage{bm}

\begin{document}

\preprint{APS/123-QED}

\title{Holographic dark energy in Brans-Dicke theory with logarithmic form of scalar field}

\author{C. P. Singh}
\email[]{cpsphd@radiffmail.com}
\author{Pankaj Kumar}
\email[]{pankaj.11dtu@gmail.com}
\affiliation{Department of Applied Mathematics,\\ Delhi Technological University, Delhi-110042, India}
\date{\today}
\begin{abstract}
 In this paper, an interacting holographic dark energy model with Hubble horizon as an infra-red cut-off is considered in the framework of Brans-Dicke theory. We propose a logarithmic form $\phi \propto ln(\alpha+\beta a)$ of the Brans-Dicke scalar field to alleviate the problems of interacting holographic dark energy models in Brans-Dicke theory. We find that the equation of state parameter $w_h$ and deceleration parameter $q$ are negative in the early time which shows the early time inflation. During the evolution the sign of parameter $q$ changes from negative to positive which means that the Universe expands with decelerated rate whereas the sign of $w_h$ may change or remain negative throughout the evolution depending on the values of parameters. It is also observed that $w_h$ may cross the phantom divide line in the late time evolution.  The sign of $q$ changes from positive to negative during late time of evolution which explains the late time accelerated expansion of the Universe. Thus, we present a unified model of holographic dark energy which explains the early time acceleration (inflation), medieval time deceleration and late time acceleration.  We also discuss the cosmic coincidence problem. We obtain a time-varying density ratio of holographic dark energy to dark matter which is a constant of order one ($r\sim \mathcal{O}(1)$) during early and late time evolution. Therefore, our model successfully resolves the cosmic coincidence problem.
\end{abstract}

\maketitle{}

\section{Introduction}
One of the present mysteries of modern cosmology is the recent accelerated expansion of the Universe predicted by the observations of supernovae Ia \cite{riess} and confirmed by the observations of Cosmic Microwave Background Radiation \cite{koma}, Large-scale Structure \cite{aba}, Baryon Acoustic Oscillation \cite{per} and Planck data \cite{ade}. According to observations, the present epoch of evolution of the Universe is dominated by an exotic energy content dubbed as ``dark energy" (DE), which has negative pressure required to explain the accelerated expansion. In the literature, a variety of DE candidates are available some of which are: cosmological constant \cite{car}, quintessence \cite{cald}, phantom \cite{well}, chaplygin gas \cite{kam} etc. The standard $\Lambda$-cold dark matter ($\Lambda$CDM) model containing the cosmological constant $\Lambda$ is the most natural and successful model of DE. However, it has some shortcomings in the form of fine-tuning and cosmic coincidence problems. The other DE candidates are also not free from problems. For a review on DE and DE candidates, see \cite{cope}.\\
\indent Recently, the holographic dark energy (HDE), which possesses some significant properties of the quantum theory, has been proposed as a candidate of DE to explain the recent phase transition of the Universe. The HDE is based on the holographic principle proposed by 't Hooft \cite{hooft} and further discussed by Susskind \cite{suss} in the context of string theory. The origin of the HDE contains the more scientific approach in comparison of other DE candidates  and presents a better way to deal with the accelerated expansion. Cohen et al. \cite{cohen} have shown that the formation of black hole imposes an upper bound on the total energy of size L and should not exceed the mass of black hole of the same size. In the paper \cite{li}, the author assumed the largest infra-red (IR) cut-off to saturate the inequality imposed by black hole formation and obtained the density of HDE $\rho_{h}=3c^2 M_{p}^2 L^{-2}$, where $c^2$ is a dimensionless constant, $M_{p}$ stands for the reduced Planck mass and L denotes IR cut-off.\\
\indent The Hubble horizon is a natural candidate for IR cut-off which is also free from causality but Hsu \cite{hsu} found that it gives wrong equation of state (EoS) of DE. Later on, it was shown \cite{pavon} that if there is an interaction between two dark components of the Universe the identification of $L$ with Hubble horizon, $L=H^{-1}$, may give suitable EoS of DE. It also was shown that it necessarily implies a constant ratio of the energy densities of the two components regardless of the details of the interaction. Thus, the HDE models may also alleviate the cosmic coincidence problem which provides an advantage to HDE models over the other DE models.\\
\indent The Brans-Dicke (BD) theory proposed by Brans and Dicke in 1961 \cite{brans} is a natural extension of general relativity (GR). In this theory, the gravitational constant $G$ is replaced with a scalar field $\phi$ called BD scalar field which couples to gravity with coupling parameter $\omega$. The inflationary epoch has been studied widely in this theory \cite{math}. Recently, the BD theory has got interest to explain the accelerated expansion due to its association with the string theory and extra dimensional theories. This theory explains the recent accelerated expansion of the Universe and accommodates the observational data as well \cite{berto}. The BD theory provides a dynamical framework which is more suitable to study the HDE models as HDE belongs to the family of dynamical DE candidates. Therefore, it is quite natural to study the HDE models in the framework of BD theory. The HDE models have been studied in the framework of BD theory to explain the recent accelerated expansion and to alleviate the problems associated with the DE models like cosmic coincidence problem \cite{seta, shey, ban, lxu, sheykh, sheykhi}.\\
\indent In the literature, most of the models have been discussed in BD theory by assuming the power-law form of BD scalar field $\phi \propto a^n$, where $a$ is the scale factor and $n$ is a constant. It has been shown in the papers \cite{berman} that the assumption $\phi \propto a^n$ naturally leads to a constant deceleration parameter (DP) in BD theory irrespective of the matter content of the Universe. In the recent papers \cite{ban, lxu, sheykh, sheykhi}, authors have used the same power-law form of BD scalar field in HDE model and found a time-dependent DP. Thus, it has been observed that a constant as well as a time-dependent DP may be obtained with this power-law form of BD scalar field in HDE model. In our point of view there should not be two different values of DP with this form of BD scalar field. Therefore, it is natural to explore more options for the BD scalar field to over come from the shortcoming of this power-law form of BD scalar field. The very first purpose of this paper is to propose a suitable form of BD scalar field which gives only time-dependent DP to discuss the evolution of the Universe in the HDE models.\\
\indent  In this context, we propose a logarithmic form $\phi \propto ln(\alpha+\beta a)$ of BD scalar field, where $a$ as usual denotes the scale factor, and $\alpha$ and $\beta$ are positive constants such that $\alpha>1$. Using this assumption, we explore the cosmological consequences of our model. This form of BD scalar field is free from the constant value of DP which naturally arises in the power-law form. We successfully obtain the equation of state (EoS) of DE and the time-dependent deceleration parameter which explain the recent phase transition of the Universe. Moreover, the unification of early time acceleration (inflation) and late time acceleration has been observed including matter dominated era. In the early and late time evolution we find that the density ratio of holographic dark energy to dark matter is a constant of order one ($r\sim \mathcal{O}(1)$) which successfully solve the cosmic coincidence problem. Therefore, we have successfully alleviated the problem associated with the power-law maintaining the good features of it. Further, the early time inflation has been explained in contrast to power-law form. The long lasting cosmic coincidence problem associated with DE models have been alleviated in a well manner in the present model in comparison to existing models.\\
\indent The outline of our work is as follows. Section II presents the field equations of HDE model in BD theory by assuming the interaction between DM and HDE. In section III, we propose a logarithmic form of BD scalar field and discuss its cosmological consequences. Section IV is devoted to the cosmic coincidence problem. The summary of the results is discussed in section V.
 \section{Holographic dark energy in Brans-Dicke Theory}
\noindent The modified Einstein-Hilbert action for the BD theory is given by \cite{sheykh}
 \begin{equation}
 S=\int d^{4}x\sqrt{-g}\left[\frac{1}{2}(-\phi R+\frac{\omega}{\phi}g^{\mu\nu}\partial_{\mu}\phi\; \partial_{\nu}\phi)+\mathcal{L}_{m}\right],
\end{equation}
  where $R$ denotes the Ricci scalar curvature, $\phi=(8\pi G)^{-1}$ is a time-dependent scalar field called BD scalar field which couples with gravity, $\omega$ is a coupling parameter between scalar field and gravity called BD parameter and $\mathcal{L}_{m}$ represents the matter lagrangian density.\\
\indent We consider a homogeneous and isotropic flat Friedmann-Robertson-Walker (FRW) Universe given by the line element
\begin{equation}
ds^{2}=dt^{2}-a^{2}(t)(dx^{2}+dy^{2}+dz^{2}),
\end{equation}
where $a$ denotes the cosmic scale factor of the Universe. We assume that the Universe is filled with perfect fluid containing pressureless dark matter (DM) excluding baryonic matter and HDE.\\
\indent The variation of the action (1) with respect to the metric tensor, $g_{\mu\nu}$ for the line element (2) with the energy-momentum tensor of dust and HDE yield the following field equations.
\begin{equation}
H^2+H\frac{\dot\phi}{\phi}-\frac{\omega}{6}\frac{\dot\phi\;^2}{\phi\;^2}=\frac{\rho_{m}+\rho_{h}}{3\phi},
\end{equation}
\begin{equation}
2\frac{\ddot a }{a}+H^2+2 H\frac{\dot\phi}{\phi}+\frac{\omega}{2}\frac{\dot\phi\;^2} {\phi\;^2}+\frac{\ddot\phi}{\phi}=-\frac{p_{h}}{\phi},
\end{equation}
where $\rho_{m}$ and $\rho_{h}$ are, respectively the energy density of DM and HDE, and $p_{h}$ denotes the pressure of HDE. Here, the over dot denotes the derivative with respect to the cosmic time $t$. Many cosmologists \cite{amen} have considered the interaction between DM and DE. The recent cosmological observations \cite{oliv} also support this interaction. Considering the interaction factor $Q$ between DM and HDE, the conservation equations of DM and HDE are respectively given by \cite{ban, sheykhi}:
\begin{equation}
\dot\rho_{m}+3H\rho_{m}=Q,
\end{equation}
\begin{equation}
\dot\rho_{h}+3H(\rho_{h}+p_{h})=-Q,
\end{equation}
where $Q=\Gamma \rho_{h}$, $\Gamma$ stands for interaction rate. The sign of interaction rate $\Gamma$ is crucial and defines the direction of the energy transfer, i.e., for $\Gamma>0$, there is an energy transfer from HDE to DM, and for $\Gamma<0$, there is an energy transfer from DM to HDE. In the literature \cite{shey,sheykhi}, $\Gamma$ has been assumed to be proportional to the Hubble parameter, i.e., $\Gamma \propto H$ to maintain the interaction term $Q$ as a function of a quantity with units of inverse of time multiplied with the energy density. Therefore, let us consider $\Gamma=3b^2 H$ so that the interaction term becomes $Q=3b^2 H \rho_h$, where $b^2$ is a coupling constant. This assumption relies purely on dimension basis in the absence of a suitable theory.\\
\indent The dynamical equation for the BD scalar field $\phi$ is given by
\begin{equation}
\ddot\phi+3H\dot\phi=\frac{\rho_{m}+\rho_{h}-3p_{h}}{2\omega+3}.
\end{equation}\\
\indent Motivated by the holographic principle \cite{hooft}, Cohen et al. \cite{cohen} obtained an upper bound on the total energy of size $L$ imposed by black hole formation. Li \cite{li} assumed the largest possible IR cut-off to saturate this inequality and obtained the density of HDE, $\rho_{h}=3c^2 M_{p}^2L^{-2}$. In the framework of BD theory, the HDE density has the form $\rho_{h} =3c^2 \phi L^{-2}$, where $\phi= M_{p}^2=(8\pi G)^{-1}$ is a time dependent scalar field which couples to gravity with a coupling parameter $\omega$. In the literature, there are various forms of HDE depending on the different choices of IR cut-off ($L$) like particle horizon, future event horizon, Hubble horizon, Granda-Oliveros cut-off etc. The Hubble horizon is the most natural and viable candidate of IR cut-off because it is free from causality problem. Here, in this paper we choose Hubble horizon as an IR cut-off which gives HDE density as:
\begin{equation}
\rho_{h}=3\;c^2 \phi H^2,
\end{equation}
where $c^2$ is a dimensionless constant.
\section{Logarithmic form of BD scalar field and cosmological consequences}
\noindent  Many authors \cite{berman} have assumed that the BD scalar field $\phi$ evolves as a power-law of the scale factor $a$, i.e., $\phi\propto a^n$. They have observed that this assumption leads to a constant value of DP. The constant value can be obtained irrespective of matter content of the Universe. However, some authors \cite{ban,lxu,sheykh,sheykhi} have studied HDE model in BD theory with the same form of BD scalar field and have obtained a time-dependent DP. Now, the question is that why does the same form of BD scalar field lead to two different values of DP, constant and time-dependent in a same model?\\
 \indent In spite of several advantages of this form of BD scalar field, it seems from the above mentioned works that this may not be a suitable assumption to discuss the evolution of the Universe in HDE models. In other words, this form may not be suitable for those models where we want to study the phase transition of the Universe. Taking into consideration to this problem, we hereby propose that the BD scalar field evolves as a logarithmic function of the scale factor which is given by
 \begin{equation}
 \phi \propto ln(\alpha+\beta a),
 \end{equation}
 \noindent where $\alpha>1$ and $\beta>0$ are constants.\\
\indent In principle, the BD scalar field should evolve slowly to observe a slow variation of $G$. This logarithmic form of $\phi$ fulfills this requirement. In this process, the value of $\beta$ also plays an important role. It is worth noting that GR can be recovered for $\beta=0$. One can observe that this form does not give a constant value of deceleration parameter when we combine Eqs. (3) and (4) with (7). Thus, we have resolved the constant value problem of the power-law form. It provides an initial advantage to the logarithmic form over the power-law form of BD scalar field. Therefore, it will be interesting to investigate the role of this form in the evolution of the Universe in the framework of BD theory within the formalism of interacting HDE model.\\
From (6), we have
 \begin{equation}
\dot\rho_{h}+3H(1+w_{h})\rho_{h}=-3b^2 H \rho_{h},
\end{equation}
 where $w_{h}=p_{h}/\rho_{h}$ is the EoS parameter of the HDE. Using (8) and (9) into (10), we get
\begin{equation}
\frac{\beta a}{(\alpha+\beta a)ln(\alpha+\beta a)}+2\frac{\dot H}{H^2}+3(1+w_{h})=-3b^2.
\end{equation}
From (3) and (4), we obtain
\begin{widetext}
\begin{eqnarray}
\frac{\dot H}{H^2}&=&\frac{1}{2+\frac{\beta a}{(\alpha+\beta a)\;ln(\alpha+\beta a)}}\Big[-3-\frac{\omega \beta^2 a^2}{2(\alpha+\beta a)^2\;[ln(\alpha+\beta a)]^2}-\frac{3\beta a}{(\alpha+\beta a)\;ln(\alpha+\beta a)}\\ \nonumber
&+&\frac{\beta^2a^2}{(\alpha+\beta a)^2\; ln(\alpha+\beta a)}-\frac{3w_{h}}{1+r}\Big\{1+\frac{\beta a}{(\alpha+\beta a)\;ln(\alpha+\beta a)}-\frac{\omega \beta^2 a^2}{6(\alpha+\beta a)^2\;[ln(\alpha+\beta a)]^2}\Big\}\Big],
\end{eqnarray}
\end{widetext}
where $r=\rho_{m}/\rho_{h}$ represents the energy density ratio. Using (12) into (11), we get
\begin{widetext}
\begin{equation}
w_{h}=\frac{(r+1)\left[-3b^2\left(2+\frac{\beta a}{(\alpha+\beta a)\;ln(\alpha+\beta a)}\right)-\frac{2\beta^2a^2}{(\alpha+\beta a)^2\; ln(\alpha+\beta a)}+\frac{\beta a}{(\alpha+\beta a)\;ln(\alpha+\beta a)}+\frac{(\omega-1)\beta^2a^2}{(\alpha+\beta a)^2\; [ln(\alpha+\beta a)]^2}\right]}{3\big[2r+\frac{(r-1)\beta a}{(\alpha+\beta a)\;ln(\alpha+\beta a)}+\frac{\omega \beta^2a^2}{3(\alpha+\beta a)^2\; [ln(\alpha+\beta a)]^2}\big]}.
\end{equation}
\end{widetext}
To discuss the behavior of $w_h$, let us first discuss the terms present in (13) which have significant contribution in the behavior of $w_h$. We observe that the terms $\frac{\beta a}{(\alpha+\beta a)\;ln(\alpha+\beta a)}$ and $\frac{\beta^2a^2}{(\alpha+\beta a)^2\;ln(\alpha+\beta a)}$ are zero at $a=0$ and both the terms converse to zero as $a\rightarrow \infty$. We find that both the terms start from zero, achieve maximum value during the evolution and then converge to zero in the late time evolution. Further, we observe that both the terms attain the maximum value asymptotically for sufficiently small values of $\alpha$ and sufficiently large values of $\beta$. The maximum value of the term $\frac{\beta a}{(\alpha+\beta a)\;ln(\alpha+\beta a)}$ lies in the interval ]0, 1[ whereas the maximum value of the term $\frac{\beta^2a^2}{(\alpha+\beta a)^2\;ln(\alpha+\beta a)}$ lies in the interval ]0, 0.41[ depending on the value of $\alpha$ only. It is to be noted that the maximum value does not depend on the value of $\beta$ in either term. \\
\indent The values of $\omega$ and $b^2$ also play an important role in the value of $w_{h}$. The value of $b^2$ is expected to be a small positive constant. The value of BD parameter $\omega$ has been constrained by various astronomical and cosmological observations. The solar system experiment Cassini gave a very stringent high bound $\omega>40000$ \cite{bert} for spherically symmetric solution in the parameterized post Newtonian formalism. The solar system constraints on $\omega$ may not be consistent at the cosmological scales, therefore, the cosmological constraints are required to study the large scale properties of the Universe. In this context, Acquaviva et al. \cite{aqua} have found $\omega>120$ at 95\% confidence level, Wu and Chen \cite{wu} have found $\omega <-120$ or $\omega>97.8$ at 95.5\% confidence level whereas Li et al. \cite{Li} have obtained $\omega>181.65$ at 95\% confidence level. However, all these constraints on $\omega$ depend on the choice of a model. In the present paper, we will consider only the value $\omega>0$.\\
\indent Now, let us discuss the behavior of EoS parameter $w_{h}$ of HDE as obtained in Eq. (13). The sign of $w_{h}$ depends on the sign of numerator as the denominator has only positive values. In the beginning of the evolution ($a=0$), all the terms of the numerator are zero except the first term which has negative sign. Therefore, we observe a negative value of $w_h$ as
\begin{equation}
w_h= -b^2\left(1+\frac{1}{r}\right).
\end{equation}
Thus, HDE has EoS of DE in the beginning which is required to explain the inflation supposed to happen in very early time of the evolution. As the Universe evolves, the terms $\frac{\beta a}{(\alpha+\beta a)\;ln(\alpha+\beta a)}$ and $\frac{\beta^2a^2}{(\alpha+\beta a)^2\;ln(\alpha+\beta a)}$ attain their maximum value, therefore, the last term of numerator containing $\omega$ starts to dominate as a large value of $\omega$ has been expected from the observations. Thus, we may observe a change in the sign of EoS parameter $w_h$ from negative to positive and HDE may start behaving like radiation ($w_{h}>0$). It is evident from (13) that the first term of the numerator has a constant part, and the terms $\frac{\beta a}{(\alpha+\beta a)\;ln(\alpha+\beta a)}$ and $\frac{\beta^2a^2}{(\alpha+\beta a)^2\;ln(\alpha+\beta a)}$ converse to zero in the late time evolution. Therefore, ultimately the sign of $w_h$ changes from positive to negative. Thus, HDE achieves the EoS of DE again during the evolution which is required to explain the late time acceleration of the Universe. It is also obvious that HDE achieves dust like behavior ($w_{h}=0$) whenever $w_h$ changes sign from positive to negative or viceversa. Zimdahl and Pav\'{o}n \cite{zim} for HDE, and Zlatev et al. \cite{zla} for tracker quintessence scalar field have also observed radiation and dust like behavior in the early time evolution.\\
\begin{figure}
  \centering
  \includegraphics[width=8.5cm, height=6cm]{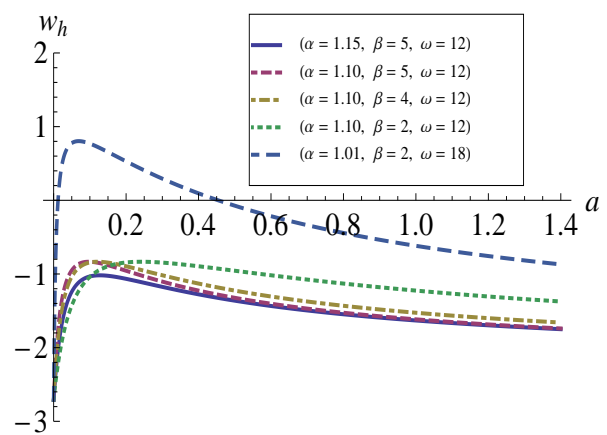}\\
  \caption{ The evolution of $w_h$ verses $a$ for different values of $\alpha$, $\beta$ and $\omega$ with $b=1.05$ and $c=0.77$.}\label{1}
\end{figure}
  \begin{figure}
  \centering
  \includegraphics[width=8.5cm, height=6cm]{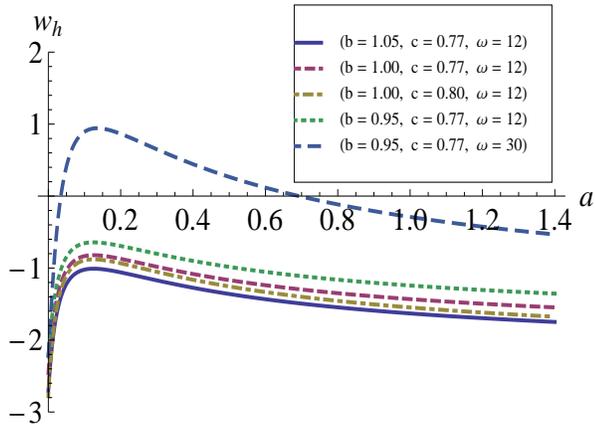}
  \caption{The evolution of $w_h$ verses $a$ for different values of $b$, $c$ and $\omega$ with $\alpha=1.15$ and $\beta=5$.}\label{2}
\end{figure}
\indent We also emphasise that for a small value of $\omega$ or a large value of $b^2$ or a suitable combination of their values, the EoS parameter $w_h$ is negative throughout the evolution. Thus, the HDE behaves like DE throughout the evolution. As it is mentioned that the maximum value of the terms $\frac{\beta a}{(\alpha+\beta a)\;ln(\alpha+\beta a)}$ and $\frac{\beta^2a^2}{(\alpha+\beta a)^2\;ln(\alpha+\beta a)}$ depend on the value of $\alpha$, the change in sign of $w_h$  also depends on the value of $\alpha$. The possible behaviors of $w_h$ are shown in Figs. 1 and 2 for different values of parameters which show the similar behaviors as discussed above. We have used value of $r$ from (21) and have assumed the present value of scale factor $a_0=1$ here and thereafter for our convenience.\\
 \indent It is worthy to note that if we take constant value of $\Gamma$ as taken in \cite{pavon}, the Eq. (13) in the late time evolution ($a\rightarrow \infty$) reduces to its respective expression in GR for HDE model, that is,
\begin{equation}
w_{h}\approx-\left(1+\frac{1}{r}\right)\frac{\Gamma}{3H}.
\end{equation}
Now, for time-dependent $\Gamma=3b^2 H$, we get the expression of $w_h$ for late time evolution as
\begin{equation}
w_{h}\approx-b^2\left(1+\frac{1}{r}\right).
\end{equation}
  One can observe from (16) that the EoS parameter converses to a negative constant value in the late time evolution. It means that as the Universe enters in the late time accelerating phase, the HDE behave like DE forever. In the absence of interaction ($b^2=0$), we obtain a positive EoS parameter of HDE ($w_h>0$) from (13) through out the evolution. It means accelerated expansion is not possible in this case as observed in the papers \cite{pavon,lxu}. It is also observed that $w_{h}$ may cross the phantom divide ($w_{h}=-1$) for a suitable value of $b^2$ in the late time evolution.\\
\indent In the present model, the dynamics of the Universe depends not only on DM and HDE, but also on the BD scalar field. It will be early to conclude about the evolution of the Universe only on the basis of EoS of HDE. Therefore, it is important to discuss the behavior of deceleration parameter, $q=-\frac{a\ddot{a}}{\dot{a}^2}$ to make a precise conclusion. We obtain the value of $q$ after dividing Eq. (4) by $H^2$, and using (8) and (9) which is given as
\begin{widetext}
\begin{equation}
q=\frac{3c^2 w_{h}+1-\frac{\beta^2a^2}{(\alpha+\beta a)^2\;ln(\alpha+\beta a)}+\frac{2\beta a}{(\alpha+\beta a)\;ln(\alpha+\beta a)}+\frac{\omega \beta^2 a^2}{2(\alpha+\beta a)^2\;[ln(\alpha+\beta a)]^2}}{2+\frac{\beta a}{(\alpha+\beta a)\;ln(\alpha+\beta a)}}
\end{equation}
\end{widetext}
Let us examine the sign of deceleration parameter to discuss the early and late time evolution. In the beginning of evolution at $a=0$, we obtain $q=\frac{3c^2 w_h+1}{2}$. Using the initial value of $w_h$ given by Eq. (14), we get $q=\frac{1}{2}\left[1-3b^2c^2\left(1+\frac{1}{r}\right)\right]$, which produces accelerated expansion for $3b^2c^2\left(1+\frac{1}{r}\right)>1$. Thus, we observe inflationary era which has been expected to happen in the beginning of the evolution to resolve the problems of Big-Bang cosmology. As $w_{h}$ may show sign change from negative to positive during the evolution and the last term containing $\omega$ in numerator of Eq. (17) starts to dominate due to large value of $\omega$, a change in the sign of deceleration parameter from negative to positive may also be observed. If $w_h$ is negative throughout the evolution then also we observe sign change in $q$ as shown in Figs. 3 and 4. Thus, we observe a decelerated expansion of the Universe after inflation.
\begin{figure}
  \centering
  \includegraphics[width=8.5cm, height=6cm]{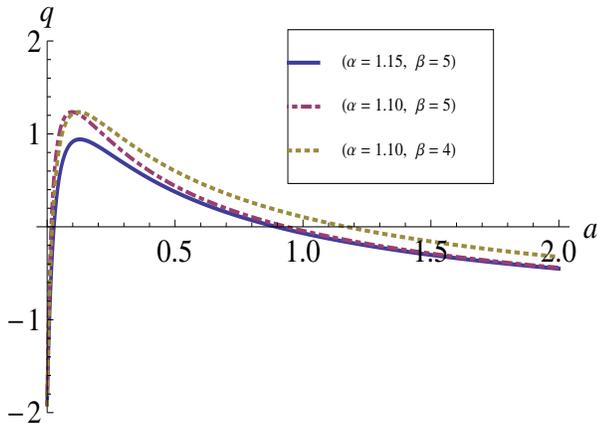}\\
  \caption{The evolution of $q$ verses $a$ for different values of $\alpha$ and $\beta$ with $b=1.05$ and $c=0.77$.}\label{3}
\end{figure}
Since the terms $\frac{\beta a}{(\alpha+\beta a)\;ln(\alpha+\beta a)}$ and $\frac{\beta^2a^2}{(\alpha+\beta a)^2\;ln(\alpha+\beta a)}$ converse to zero in the late time evolution, the late time value of $q$ is obtained as
 \begin{equation}
 q\approx\frac{3c^2 w_h+1}{2},
 \end{equation}
 Using the late time value of $w_h$ given by (16) into (18), we get
 \begin{equation}
 q\approx\frac{1}{2}\left[1-3b^2c^2\left(1+\frac{1}{r}\right)\right].
 \end{equation}
 \begin{figure}
  \centering
  \includegraphics[width=8.5cm, height=6cm]{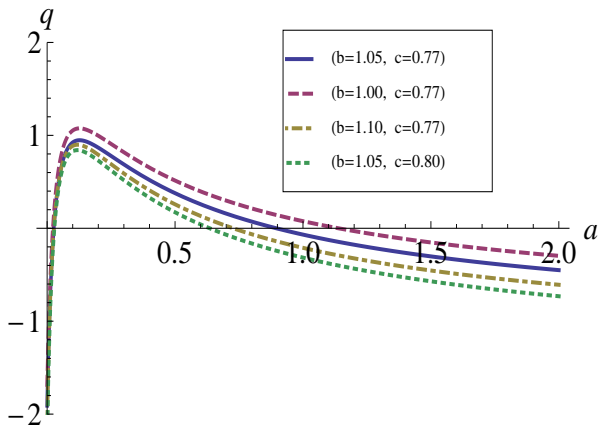}\\
  \caption{The evolution of $q$ verses $a$ for different values of $b$ and $c$ with $\alpha=1.15$ and $\beta=5$.}\label{4}
\end{figure}
  Here, $q$ will be negative if $3b^2c^2\left(1+\frac{1}{r}\right)>1$, the same as for early time. Therefore, a sign change of $q$ from positive to negative may be observed which successfully explains the late time phase transition (deceleration to acceleration) of the Universe. Indeed, HDE shows almost same behavior in the early and late time evolution of the Universe. If there is no interaction ($b^2=0$), we observe $q>0$ through out the evolution. Thus, using the value of $q$ we reconfirm that the accelerated expansion is not possible in the case of no interaction. It can be seen that it is not only for $b^2=0$, but also for a small value of $b^2<<1$, a decelerating Universe may be observed throughout the evolution. The other parameters of the model will not change this scenario. Further, we observe that a sufficiently large value of $b^2$ is able to accelerate the expansion through out the evolution. Therefore, the value of coupling parameter $b^2$ plays an important role to observe the evolution of the Universe which is consistent with the observations. It means that a suitable interaction between the dark components of the Universe, i.e., a suitable value of $b^2$ is required to be consistent with the observations.\\
   \indent After having a suitable interaction, the most important parameters are $\alpha$ and $\beta$, which are due to the logarithmic form of BD scalar field. The terms containing $\alpha$ and $\beta$ in Eq. (17) start from zero, evolve up to maximum value and converse to zero in the late time which lead to observe two phase transitions of the Universe, i.e., the early time inflation and the late time acceleration. It is worthy to mention here that for the power-law form of BD scalar field, only a late time acceleration is possible. The values of $\alpha$ and $\beta$ are important to decide the time when the phase transitions, inflation and late time acceleration, will happen which can be seen in Figs. 3 and 4. Thus, to accommodate the observed history of the Universe the value of $\alpha$ and $\beta$ are significant. It is also evident from (17) that for a large value of $\omega$ the inflation ends early and the late time acceleration occurs late, and viceversa. So far we have observed that a number of combination of the parameters are possible to explain the whole history of the Universe but let observations to decide the best combination of the parameters.\\
  \indent We have shown that the present model of dark energy is able to explain the whole known history of the Universe starting from inflation to recent accelerated expansion of the Universe including the matter dominated era (radiation+dust). Thus, this model presents a unified model of interacting HDE in the framework of BD theory with logarithmic form of BD scalar field. We emphasise that the logarithmic form proposed in this paper has played an important role to explain the evolution of the Universe in more better way. In the next section, we will show that this form also plays a significant role to resolve the coincidence problem.
  \section{Cosmic Coincidence Problem}
\noindent Let us investigate the present model on the ground of cosmic coincidence problem \cite{zla,stein}. Using the conservation Eqs. (5) and (6), the evolution of energy density ratio $r$ can be obtained as
\begin {equation}
\dot r=3Hr\left[w_{h}+\left(\frac{1+r}{r}\right)b^2\right],
\end{equation}
which is the same expression as obtained in \cite{pavon,ban} except we have $b^2$ at the place of $\Gamma/3H$ because they have chosen $\Gamma$ as a constant whereas we have taken $\Gamma=3b^2 H$. The nature of evolution of $r$ in our model is different from the mentioned works as we have taken time-dependent $\Gamma$ and it also depends on the value of $w_{h}$. Using the late time value of $w_{h}$ given by (16) into (20) we obtain $\dot r=0$, i.e., in the late time evolution $r$ has a constant value which is a significant feature to solve the coincidence problem. Pav\'{o}n and Zimdahl \cite{pavon} have obtained a constant value of $r$ in GR, however, $r$ is expected to be a time-varying value. Therefore, the authors assumed a time-varying value of $c$ to obtain a time-varying $r$. It has also been suggested to replace the Hubble horizon by the future event horizon to achieve a time-varying value of $r$ \cite{wang}. Although Banerjee and Pav\'{o}n \cite{ban} have obtained time-dependent value of $r$ in Brans-Dicke theory but they achieved a soft coincidence only. They argued that $r$ can vary more slowly in their model than in the conventional $\Lambda$CDM model. It have been demonstrated by Zhang et al. \cite{zhang} that the interacting chaplygin gas model has a better chance to solve the coincidence problem in comparison of interacting quintessence and interacting phanton models. In the literature, many proposals have been made to solve the coincidence problem \cite{campo,velten} but the problem still exists.\\
\indent Let us find the value of $r$ for our model to analyse coincidence problem in more detail. From (3), the value of $r$ is obtained as
\begin{widetext}
\begin {equation}
r=-1+\frac{1}{c^2}+\frac{\beta a}{c^2(\alpha+\beta a)\;ln(\alpha+\beta a)}-\frac{\omega\beta^2 a^2}{6c^2(\alpha+\beta a)^2\;[ln(\alpha+\beta a)]^2}.
\end{equation}
\end{widetext}
Here, we obtain a time-dependent value of $r$ which has a constant and finite value $r=-1+\frac{1}{c^2}$ at the beginning of the evolution as the last two terms are zero at $a=0$. In the late time evolution, we also obtain a constant and finite value $r\approx-1+\frac{1}{c^2}$ as the last two terms converge to zero. The same constant value of $r$ has been obtained through out the evolution in GR \cite{pavon} but we have obtained a time-dependent value of $r$ which has constant values in the early and late time evolution. Here, it will be interesting to quote a paper of Campo et al. \cite{campo} ``Obviously, a mechanism that makes $r$ tends to a constant today or decrease its rate to a lower value than the scale factor expansion rate ameliorates the coincidence problem significantly, but it does not solve it in full. To do so the said mechanism must also achieve $r_{0}\sim \mathcal{O}(1)$". We have obtained $r_{0}\sim -1+\frac{1}{c^2}$ which excellently satisfies the requirement $r_{0}\sim \mathcal{O}(1)$ as most of the observational constraints on $c$ obtain $0.5<c<1$ \cite{Xu,mli,lixin}. In the other words, the cosmic coincidence problem has been resolve completely. In fact, we observe $r\sim \mathcal{O}(1)$ in the early time as well as in the late time evolution in our model. Therefore, the cosmic coincidence problem does not seem a problem because it is not a coincidence that we are living in a time where $r\sim \mathcal{O}(1)$, it have been observed in the early time also. Thus, in conclusion we can say that the cosmic coincidence problem has been resolved in a significantly well manner in the present model.\\
\indent From (16) and (20), it is clear that whatever the value of $b^2$ be chosen, we obtain $\dot r=0$ in the late time evolution. It means the coupling parameter $b^2$ between DM and HDE does not play a significant role in the alleviation of coincidence problem. This is also evident from the value of $r$ given by (21) which has no $b^2$ term. Also, one may observe that the parameter $c$ is not significant to resolve the coincidence problem. Although, it plays an important in the value of $r$ at any given time. Actually, the last two terms in Eq. (21) which come due to the assumption of logarithmic form of BD scalar field,  play an important role to decide the coincidence problem.  These two terms converse to zero in the late time and may vary sufficiently slow at present due to which $r$ converses to a constant value in the late time and vary sufficiently slow at present. The suitable values of the parameters $\alpha$ and $\beta$  ensure the slow variation of $r$ at present time. The variation of $r$ is shown in the Fig. (5) which clearly verify our claim.\\
 \begin{figure}
  \centering
  \includegraphics[width=8.5cm, height=6cm]{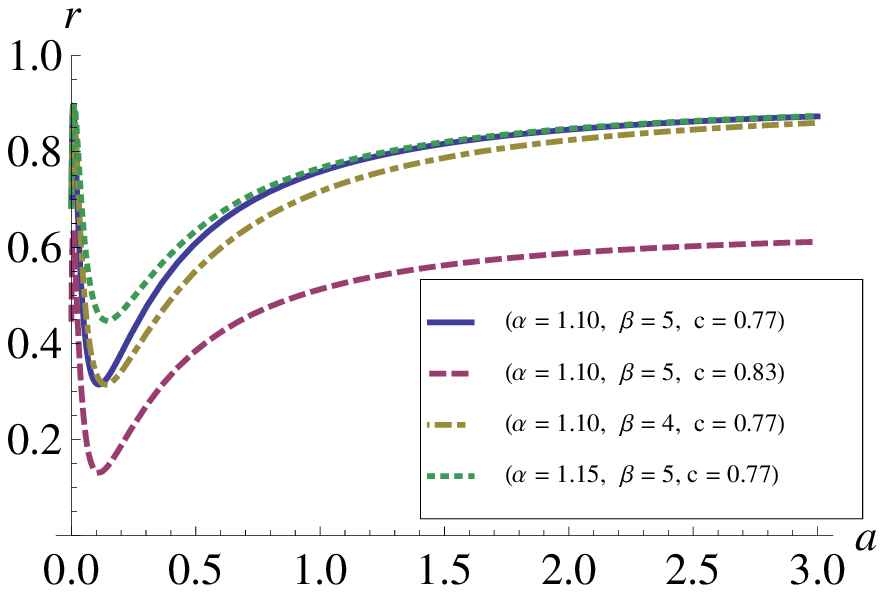}\\
  \caption{ The graph of $r$ verses $a$ for different values of $\alpha$, $\beta$ and $c$ with $\omega=12$.}\label{3}
\end{figure}
 \indent Let us find the value of $c$ to match with the observational values using the observed value of $r$. According to recent observations, the present value of $r$ is $\approx$ $\frac{3}{7}$, where matter has $\approx$ $30\%$ and DE has $\approx$ $70\%$ of total energy content of the Universe. We obtain the value $c\approx 0.8367$ using the theoretical relation $r\approx -1+\frac{1}{c^2}$ from (21) for late time evolution. Xu at al. \cite{Xu} have obtained similar value $c=0.807^{+0.165}_{-0.160}$ for the HDE model in the framework of Brans-Dicke theory. Using the observations from various probes like WMAP, BAO, Planck etc. Li et al. \cite{mli} have obtained values of $c$ which also show similarity with the above values, however, all these models are different.\\
\section{Conclusion}
\indent In this paper, we have studied interacting HDE model with Hubble horizon as an IR cut-off in the framework of BD theory. We have pointed out a serious problem associated with the assumption of power-law form ($\phi\propto a^n$) of the BD scalar field. It has been observed that this form of BD scalar field gives constant and time-dependent deceleration parameter for same model. Therefore, taking into consideration of this problem, we have proposed a logarithmic form of BD scalar field which always gives a time-varying deceleration parameter. We have observed that this form of BD scalar field is consistent with the cosmic evolution of the Universe. We have obtained EoS parameter and deceleration parameter to discuss the early and late time evolution of the Universe. We have also discussed the cosmic coincident problem. A summary of the main findings is as follows.\\
\indent In the early time evolution, we observe $w_h<0$ and $q<0$ which explains the inflationary era of the cosmic evolution supposed to happened in the very early Universe to resolve the problems of Big-Bang cosmology. Further, the sign change of $q$ from negative to positive has been observed. Thus, we observe the matter dominated era of the evolution. In the late time evolution, we have observed another sign change of $q$ from positive to negative which explain the late time acceleration of the Universe. Thus, this model presents a unified model of interacting HDE in the framework of BD theory with logarithmic form of BD scalar field. If we consider $\Gamma$ as a constant then the EoS parameter $w_h$ reduces to the corresponding form of GR in the late time. The EoS parameter converges to a constant value and for a suitable value of $b^2$ it may cross the phantom-divide line $w_h=-1$ in the late time evolution. If there is no interaction between HDE and DM, i.e., $b^2=0$, it has been observed that the EoS parameter $w_h$ and deceleration parameter $q$ are always positive, therefore, the accelerated expansion is not possible in this case.\\
\indent We have also discussed the cosmic coincidence problem, a long standing problem with DE models, which has been resolved effectively in this paper. We have observed a time-varying energy density ratio $r$ which has a constant value in early and late time evolution of the Universe. The present value $r_{0}$ satisfies the condition $r_{0}\sim \mathcal{O}(1)$ which is consistent with the observations. Using the present value of $r$ obtained by observations, we get theoretical value $c=0.8367$ which shows similarity with the observed values of $c$.\\
\indent In conclusion, the advantage of our model is that we have alleviated the problem associated with the power-law containing the good features of it. In addition, our model presents a unification of early time inflation and late time acceleration including the matter dominated era which have not been obtained through power-law form. In the present model, the cosmic coincidence problem has been alleviated more effectively in comparison of existing models.

\end{document}